\documentclass{vgtc}                          

\graphicspath{{figures/}{pictures/}{images/}{./}} 

\usepackage{times}                     

\usepackage{tabu}                      
\usepackage{booktabs}                  
\usepackage{lipsum}                    
\usepackage{mwe}                       

\usepackage{mathptmx}                  

\usepackage{graphicx}
\usepackage{subcaption}

\onlineid{0}

\vgtccategory{Research}

\vgtcinsertpkg

\title{Towards Extended Reality Intelligence for Monitoring and Predicting Patient Readmission Risks}

\author{Martin Sanchez
\and Nick Tran
\and Vuthea Chheang\thanks{e-mail:vuthea.chheang@sjsu.edu}}
\affiliation{\scriptsize Department of Computer Science, San Jose State University, USA}

\teaser{
  \centering
  \includegraphics[width=\linewidth]{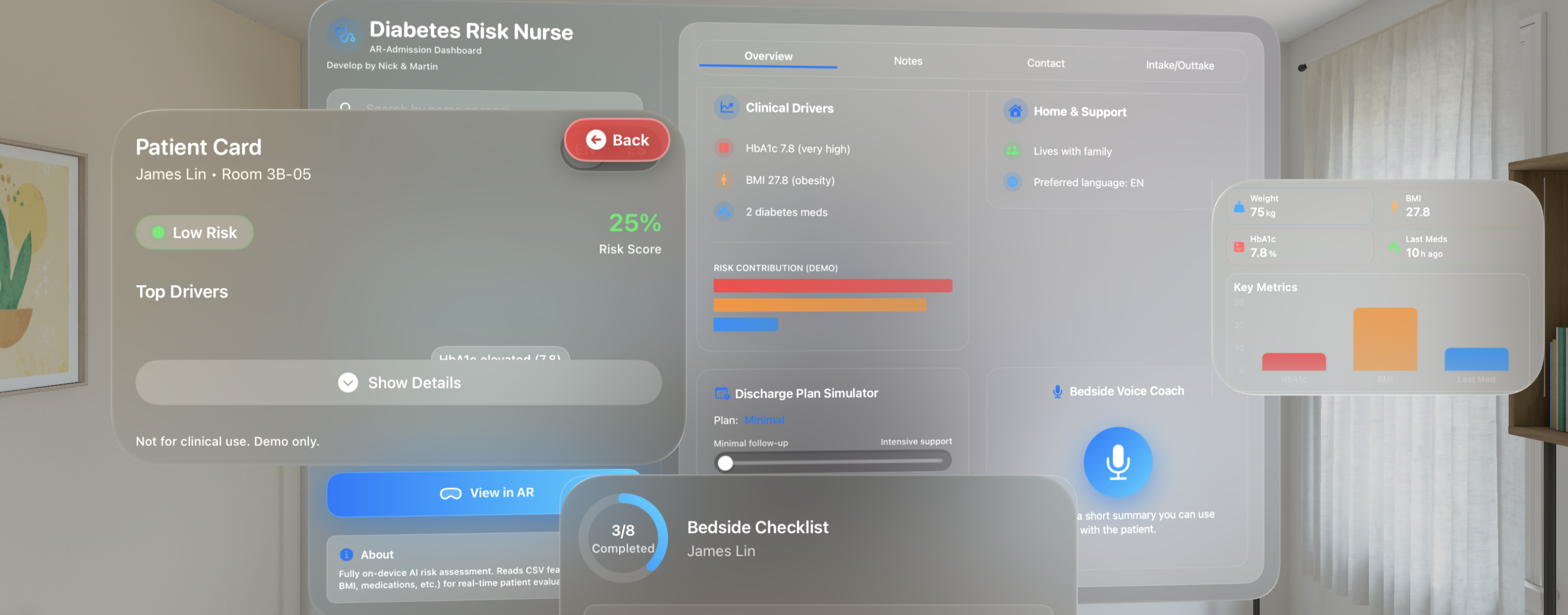}
  \caption{Extended Reality Intelligence (XRI) interfaces for monitoring and predicting diabetic patient readmission risks.}
  \label{fig:teaser}
}

\abstract{
    Hospital readmissions remain a challenge for healthcare systems, especially among patients with chronic conditions such as diabetes. Unplanned readmissions within 30 days are costly, strain hospital resources, and can indicate poor care coordination or discharge planning. In this work, we explore the use of machine learning to predict readmission risk for diabetic inpatients and propose a mixed reality (MR) to provide effective visualization and insights.
    We trained an XGBoost classifier after data cleaning, encoding, and feature engineering. The model achieved an Area Under the Receiver Operating characteristic Curve (AUROC) of 0.72 and an Area Under the Precision-Recall Curve (AUPRC) of 0.11. Key predictive factors included prior inpatient visits, discharge disposition, and glycemic control indicators such as A1C (blood sugar test) results and medication adjustments.  
    Additionally, we developed an MR prototype that visualize patient records and predictions containing risk level, major contributing factors, and a concise summary of care. Together, the predictive model and the MR interface aim to improve clinician awareness and communication around readmission risk in real-time clinical settings.
} 


\begin{document}


\firstsection{Introduction}

\maketitle

Hospital readmission is a critical metric of healthcare quality and efficiency~\cite{zuckerman2016readmissions}. 
Each year, avoidable readmissions cost hospitals billions of dollars and lead to increased patient stress and resource strain. For patients with diabetes, a population prone to complications and frequent admissions, predicting who is at risk of being readmitted within 30 days is especially important.
Traditional tools such as the Length of stay, Acuity of admission, Comorbidities, and Emergency department visits (LACE) index and the hospital score provide quick estimates but rely on limited linear relationships and are often buried deep within electronic health record (EHR) dashboards. Additionally, these systems might not always accessible during bedside consultations, where timely communication is most important. 

Previous studies have examined hospital readmissions from both clinical and computational perspectives~\cite{artetxe2018predictive, mosadeghi2016feasibility, ridout2021effectiveness}.   
Dhaliwal and Dang \cite{dhaliwal2025readmissions} emphasized that reducing 30-day readmissions is a key national quality priority, noting that early risk identification can significantly lower costs and improve outcomes. Machine-learning approaches have gained traction in recent years: Sharma \textit{et al.} \cite{sharma2022mlreadmissions} demonstrated that gradient-boosted models outperform traditional regression in predicting heart-failure readmissions, achieving modest but meaningful improvements in AUROC.
Robinson \textit{et al.} \cite{robinson2019predictors} found that the number of discharge medications alone was a strong indicator of readmission risk, performing comparably to the HOSPITAL and LACE indices. These findings suggest that both data richness and feature selection critically influence performance.
Despite these advances, current predictive tools remain limited in usability and required manually navigation in EHR dashboards that require the clinicians to manually navigate~\cite{hwang2023toward, hirzle2023xr, reiners2021combination}. 

Extended reality (XR) shows potential for mitigating readmission risks as it transforms complex, buried predictive data into intuitive and spatial visualizations~\cite{ridout2021effectiveness, chheang2024towards, chheang2024advanced}.
To address challenges in clinical decision-making and the need for effective visualization, we combine a tabular machine learning (ML) model with an immersive visualization layer. Specifically, we investigate whether displaying readmission risk through a mixed-reality (MR) \textit{``PatientCard"} can make predictive information more immediate and interpretable at the point of care (see~\autoref{fig:teaser}).
We explore machine-learning models based on \textbf{Extreme Gradient Boosting (XGBoost)} classifier to predict 30-day readmission risk for diabetic inpatients and propose an MR prototype to effectively visualize the information and predictions. 
The results show model performances on a large de-identified hospital dataset and an integration of its prediction results into an MR application that displays patient-specific risk information in a concise and intuitive format.


\section{Materials and Methods}

\subsection{Dataset}
We use the dataset \textit{Diabetes 130-US Hospitals for Years 1999--2008}~\cite{ strack2014impact} as a use case. 
The dataset contains roughly $100,000$ hospital encounters collected across 130 U.S. hospitals and integrated delivery networks. Each record represents a single hospital admission with 50 attributes describing patient demographics, diagnoses (ICD-9 codes), laboratory results, medications, and prior utilization history. The target variable is whether the patient was readmitted within 30 days of discharge, encoded as binary (\texttt{<30 = 1}, \texttt{>30/No = 0}).

\subsection{Data Preprocessing}

The target variable was converted to binary.
The dataset was first split into $80\%$ training pool and $20\%$ test set using stratified sampling to preserve the readmitted vs. non-readmitted ratio. The $80\%$ training pool was then further split into $64\%$ training and $16\%$ validation (an 80/20 split of the training pool), resulting in an overall 64/16/20 train–validation–test split ($random_state = 42$). Because the positive class (readmitted within 30 days) represents only about $5\%$ of all records, class imbalance was addressed using the \texttt{$scale\_pos\_weight$} parameter in XGBoost, set to approximately the ratio of negatives to positives ($\approx 15{:}1$).


The engineered features reflect factors identified in the literature as major contributors to hospital readmission risk, such as prior utilization, discharge planning, and chronic disease control.

\subsection{Model Selection}
We used an XGBoost classifier, chosen for its robustness on tabular healthcare data and ability to handle nonlinear relationships and missing values. Compared to linear or logistic models, XGBoost can naturally model feature interactions and capture threshold effects (e.g., increased risk once inpatient visits exceed a certain count). The model was implemented in Python (3.12.2) using XGBoost 3.0.5.
The hyperparameters were tuned with small grid sweeps in the validation set. The final configuration included: \texttt{max\_depth = 6}, \texttt{learning\_rate = 0.05}, \texttt{n\_estimators = 600}, \texttt{subsample = 0.9}, \texttt{colsample\_bytree = 0.9}, and \texttt{min\_child\_weight = 1}. Early stopping used 50 rounds based on validation AUPRC.

\subsection{Model training}
The XGBoost classifier was trained using the training and validation sets. We enabled early stopping with 50 rounds: if the validation AUCPR did not improve for 50 consecutive boosting iterations, training terminated and the best iteration was retained. The validation set (16\% of data) was used exclusively for model selection and hyperparameter tuning, while the 20\% test set was used only for final evaluation.

\subsection{Evaluation Metrics}
Model performance was assessed using multiple metrics to account for class imbalance. The AUROC measured overall ranking ability, while the AUPRC provided a more informative metric for rare events. Additional metrics included precision, recall, F1-score, balanced accuracy, and the confusion matrix at a decision threshold of 0.5. All evaluations were conducted on the test set.

\subsection{Mixed Reality (MR) Implementation}
To visualize predictions in a clinician-friendly way, the model's outputs were integrated into an MR prototype developed with Apple's Vision Pro headset. The prototype was built in XCode 15 using SwiftUI, RealityKit, and ARKit frameworks. It displays a floating ``PatientCard'' at the bedside showing a color-coded risk indicator (green = low, yellow = medium, red = high), top contributing features, and a short plain-language care plan. The interface runs entirely on-device using Apple's Core ML for model inference, ensuring privacy and low latency.

To maximize MR reliability across rooms, we use a \textbf{head-anchored} card by default: the card is attached to a head anchor in RealityKit and positioned slightly below eye line at $\sim$0.6\,m for comfort. The panel is billboarded so it always faces the user. This eliminates failure cases due to inconsistent plane detection in unfamiliar environments. In addition, adding \textbf{plane anchoring} (desk/bed) with automatic fallback is essential to head-anchor when surfaces are not confidently detected.

\section{Analysis and Results}

\subsection{Exploratory Data Analysis}
Preliminary analysis revealed a strong class imbalance, with only about 5\% of encounters resulting in a 30-day readmission. This confirmed the need for imbalance handling in the modeling stage.

\begin{figure}[t]
\centerline{\includegraphics[width=0.45\textwidth]{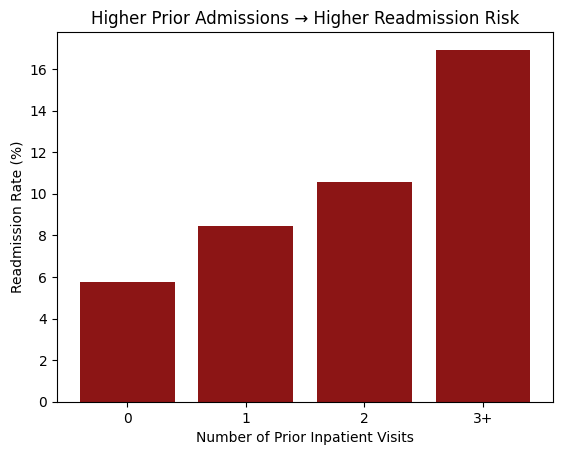}}
\caption{Readmission rate by number of prior inpatient visits. Patients with $\geq$3 prior visits had nearly triple the readmission risk.}
\label{fig:prior_inpatient}
\end{figure}

\begin{figure}[t]
\centerline{\includegraphics[width=0.45\textwidth]{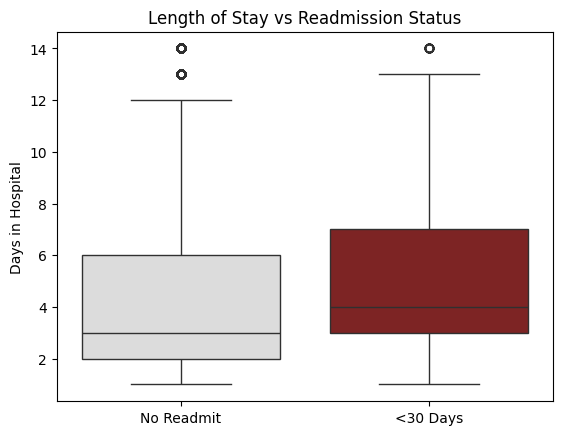}}
\caption{Length of stay (LOS) vs.\ readmission status. Readmitted patients had longer median stays, indicating higher illness severity.}
\label{fig:los_boxplot}
\end{figure}

As shown in~\autoref{fig:prior_inpatient}, readmission likelihood increases sharply with the number of prior inpatient visits. Patients with three or more previous hospitalizations were nearly three times more likely to be readmitted compared to those with no prior inpatient history.
The median length of hospital stay was longer for readmitted patients (approximately four days) than for non-readmitted patients (roughly three days) (see~\autoref{fig:los_boxplot}~).

\begin{figure*}[h]
    \centering
    \begin{subfigure}[b]{\columnwidth}
        \centering
        \includegraphics[width=\textwidth]{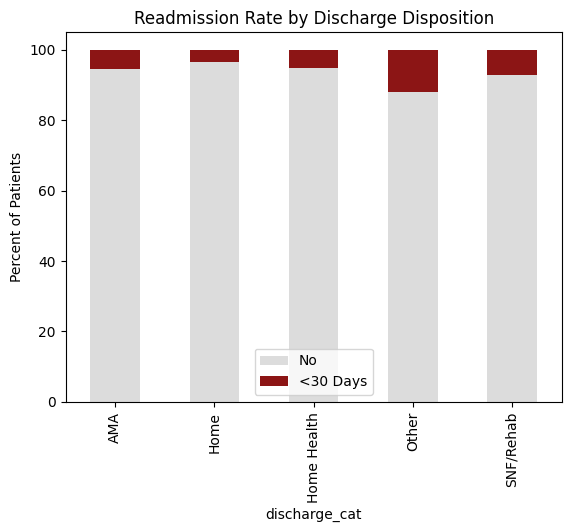}
        \caption{}
        \label{fig:discharge_bar}
    \end{subfigure}
    \hfill
    \begin{subfigure}[b]{\columnwidth}
        \centering
        \includegraphics[width=\textwidth]{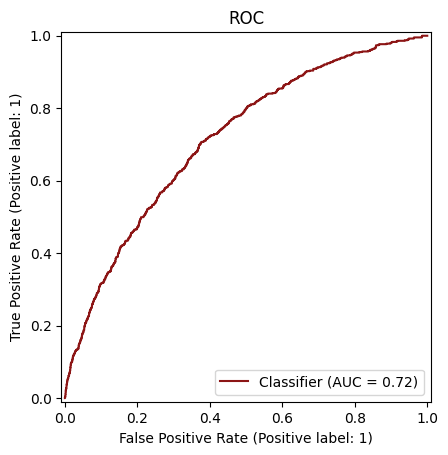}
        \caption{}
        \label{fig:roc_pr}
    \end{subfigure}
    \caption{Results of discharge disposition and model performance: (a) Readmission rate by discharge disposition -- patients discharged to rehab or AMA had the highest risk and (b) receiver operating characteristic (ROC) and precision--recall (PR) curves for the XGBoost classifier. AUROC = 0.72, AUPRC = 0.11.}
    \label{fig:image2}
\end{figure*}

Results show that discharge disposition significantly impacts readmission risk. Patients discharged to skilled nursing or rehabilitation facilities, or those who left against medical advice (AMA), were readmitted more frequently than those discharged home (see~\autoref{fig:discharge_bar}).
These results align with clinical intuition--patients with longer hospital stays, complex recovery needs, or inadequate post-discharge support are more prone to early readmission.

\subsection{Model Performance}
\autoref{fig:roc_pr} presents the ROC and precision--recall curves.
The XGBoost model achieved a test AUROC of 0.72 and AUPRC of 0.11, indicating reasonable discrimination and ranking performance for this imbalanced dataset. At a default threshold of 0.5, the model reached a recall of 0.59 and precision of 0.09, with a balanced accuracy of approximately 0.65. 


\subsection{Feature Importance}
Table~\ref{tab:feature_importance} shows the top ten features ranked by XGBoost gain score. The most influential variables were related to prior utilization (\texttt{inpatient\_ge\_2}, \texttt{number\_inpatient}, \texttt{prior\_util\_sum}), discharge planning (\texttt{discharge\_disposition\_id}, \texttt{discharge\_home}, \texttt{discharge\_snf\_rehab}), and glycemic control (\texttt{a1c\_high}, \texttt{any\_diabetes\_med}). Demographic factors such as age also contributed notably. These results reinforce prior research that frequent hospitalizations, complex discharges, and poor disease control are strong predictors of early readmission.

\begin{table}[!htbp]
\caption{Top 10 Features by XGBoost Gain}
\begin{center}
\begin{tabular}{|l|c|}
\hline
\textbf{Feature} & \textbf{Gain Score} \\
\hline
inpatient\_ge\_2 & 1302.01 \\
number\_inpatient & 253.71 \\
discharge\_disposition\_id & 152.67 \\
prior\_util\_sum & 138.62 \\
discharge\_home & 135.80 \\
discharge\_snf\_rehab & 128.60 \\
age\_mid & 126.24 \\
admission\_type\_id & 109.88 \\
any\_diabetes\_med & 104.59 \\
gender\_Male & 99.25 \\
\hline
\end{tabular}
\label{tab:feature_importance}
\end{center}
\end{table}

\begin{figure}[t]
\centerline{\includegraphics[width=0.45\textwidth]{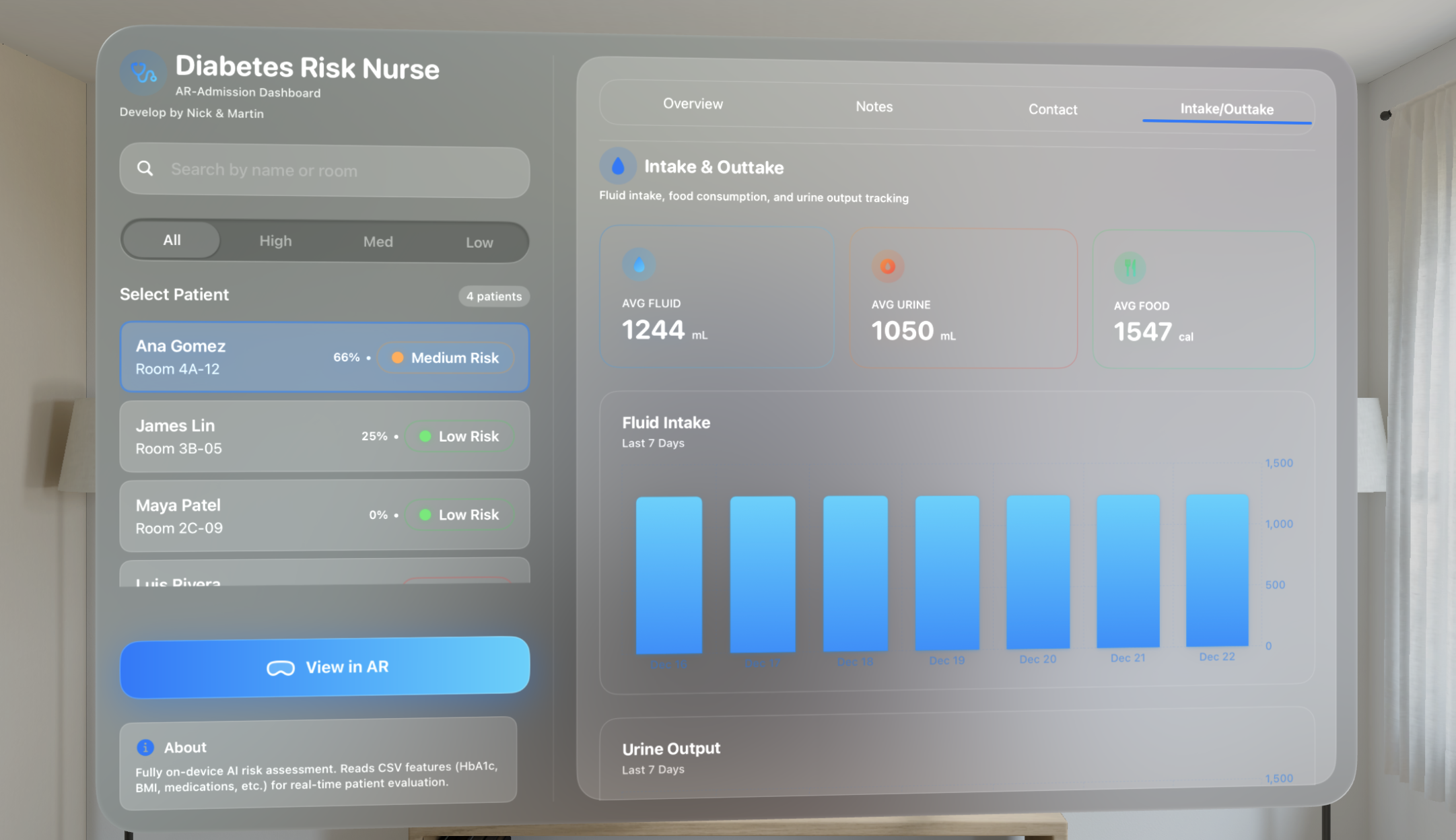}}
\caption{A ``PatientCard" XR interface showing readmission risk and top predictive factors.}
\label{fig:ar_mockup}
\end{figure}

\subsection{Extended Reality Intelligence}
To bridge the gap between data science and clinical practice, we integrate predictive modeling with a prototype MR interface (see~\autoref{fig:ar_mockup}). The goal is to make readmission risk information immediately accessible and visually intuitive for clinicians at the bedside. Instead of relying on traditional dashboards, our system displays a ``PatientCard" that summarizes each patient's 30-day readmission risk, contributing factors, and next-step recommendations in plain language.


The application reads model predictions and metadata from a JSON file or Core ML model, ensuring all computation and patient data remain on-device.
The visualization pipeline is designed as: (1) Input -- risk score and top five contributing features from the trained XGBoost model. (2) Processing-- Core ML converts the trained model into an Apple-optimized format for local inference.
(3) Output -- an interactive ``PatientCard" rendered as a floating 3D panel within the clinician's field of view.

The \textit{PatientCard} appears approximately 0.6 meters in front of the clinician, anchored at head position for reliability and comfort. The interface includes a \textit{color-coded risk indicator} (low, medium, high), a short list of the top model features contributing to the patient's risk, a plain-language summary translating technical variables into easy explanations, and basic touchless interactions---gaze to select, pinch gesture to expand details.
The UI design emphasizes readability, stability, and hygiene in clinical environments. All text and icons use high-contrast layouts and are anchored in world space with billboarding for consistent orientation.


\section{Discussion}

The results show that a ML model trained on EHR data can moderately predict 30-day readmission risk among diabetic inpatients. The model's AUROC of \textbf{0.72} and AUPRC of \textbf{0.11} exceed typical logistic or rule-based scores.
Key drivers, including prior hospital utilization, discharge disposition, and glycemic control, are clinically intuitive and align with prior literature \cite{dhaliwal2025readmissions, robinson2019predictors, sharma2022mlreadmissions}.
Our results demonstrate that ML models can extract stable and generalizable risk patterns, even when trained on historical public datasets. This suggests that the underlying physiological or causal mechanisms of risk remain consistent over time, validating the utility of legacy data for initial model development.
However, we acknowledge that medical practices and patient demographics evolve. Consequently, validation against a contemporary clinical dataset is essential to confirm the model's relevance to current healthcare environments and to uncover potentially new risk factors not present in older repositories. 
To address this, our future work focuses on establishing a partnership with a clinical institution. While this pathway presents logistical challenges -- specifically regarding data privacy regulations and ethical review timelines, securing access to longitudinal, real-world clinical data remains a priority for the next phase of this work.

While our focus is on the deployability of the ML model on the standalone XR hardware without reliance on cloud processing, which is crucial for hospital data privacy, 
the model performance in our work based on evaluation metrics is comparable to existing literature.  
Thus, the limitation lies in the model training setup. 
We relied on a single train--validation split with early stopping based on validation AUCPR. While this helps prevent overfitting, it makes performance sensitive to the particular partitioning of the data and constrains the depth of hyperparameter tuning. Future work could incorporate more robust training strategies such as $k$-fold cross-validation, Bayesian or grid search optimization, and repeated experiments across multiple random seeds to improve generalizability and stability.

Beyond model accuracy, we emphasize \textbf{clinical accessibility and usability}. Rather than relying on small widgets obscured within an EHR interface, our MR prototype surfaces the most relevant information in a glanceable ``PatientCard''. This design aims to reduce cognitive load and improve communication. While currently a proof-of-concept, our design principles and early feedback suggest that MR-based decision support enhances comprehension and workflow.
The presented prototype functions as a proof-of-concept demonstrating technical feasibility. We acknowledge that rigorous user studies and clinical validation are required to evolve this work from a prototype into a clinical settings. 
Our future work will focus on extensive evaluations in clinical environments--specifically comparing MR against traditional 2D dashboards, to quantify the system's impact on decision-making speed, user experience, and patient outcomes.

While ML models for readmission prediction are becoming increasing robust, their clinical usability is often relied on static dashboards located at a fixed station, physically disconnected from the point of care.  
XR intelligence offers a transformative approach to reducing patient readmissions by bridging the critical gap between predictive accuracy and clinical usability. We argue that high readmission rates are often driven by cognitive overload and information fragmentation in traditional EHRs, rather than a lack of data. By synergizing robust ML models, which uncover stable risk patterns, with the spatial computing capabilities, our system surfaces complex insights into a glanceable, hands-free ``PatientCard". 
This includes the mitigation of the split-attention effect. Therefore, by spatially anchoring the ``PatientCard'' and ML-driven risk indicators directly alongside the patient, the system can place data within the immediate field of view. This alignment could aim to reduce the cognitive load associated with context switching, allowing for a more seamless synthesis of digital data and physical observation.
Moreover, rather than navigating the EHR interfaces on desktop-based system to find the risk score, the clinician is presented with an immediate, top-level visual summary upon entering the room with hand-free interaction. This mechanism for data delivery is intended to accelerate triage decisions during rapid rounds.
The development of this work utilized the Apple Vision Pro. While the high cost of this hardware currently limits broad accessibility, it provides the necessary technical specifications for a viable medical interface. Specifically, the device's high-resolution display mimics the clarity of physical monitors, and its advanced spatial computing enables intuitive, hands-free interaction—a critical requirement for infection control and multitasking in hospital environments~\cite{khandekar2025augmented, wei2025scoping}. As MR hardware costs decrease over time, we anticipate these capabilities will become more accessible for general clinical use.


\section{Conclusion}
We present a proof-of-concept ``XR intelligence" system designed to enhance bedside decision-making by integrating ML models with MR.
We implemented a data-driven approach to predicting 30-day readmissions for diabetic inpatients and explored delivery of those insights using the \textit{Diabetes 130-US Hospitals} dataset as a use case.
The system visualizes patient data and risk probabilities as a spatial ``PatientCard," effectively merging digital analytics with the physical care environment. 
By providing a hands-free, glanceable interface, our proposed system aims to mitigate the split-attention effect and maintain sterility in clinical workflows. This work demonstrates the technical feasibility of the architecture and lays the foundation for future clinical usability studies.

\acknowledgments{
This work was supported by the Department of Computer Science at San Jose State University through seed funding for conference registration.
}

\bibliographystyle{abbrv}

\bibliography{references}
\end{document}